\documentclass[12pt]{iopart}
\usepackage{graphics}

\begin{document}

\title[Sinusoidally Modulated Vacuum Rabi Oscillation...]{Sinusoidally
Modulated Vacuum Rabi Oscillation of a Two-Level Atom in an Optomechanical
Cavity}
\author{Zhenshan Yang$^*$, Chenglin Bai, Xiangguo Meng, and Minghong Wang}
\address{Shandong Provincial Key Laboratory of Optical Communication Science and
Technology, School of Physical Science and Information Engineering,
Liaocheng University, Liaocheng, Shandong 252000, China}

\ead{yangzhenshan@lcu.edu.cn} \vspace{10pt}

\begin{abstract}
We study the coherent dynamics of an excited two-level atom in a vacuum
optomechanical cavity and find that the original atom-cavity Rabi
oscillation is sinusoidally modulated by the light-mechanics coupling as the
Rabi splitting is on resonance with the mechanical mode. We develop an
analytic model in a three-dimensional Hilbert subspace to explain this
phenomenon and employ numerical simulations of the density-matrix master
equation to confirm our analysis. We also show that the modulated Rabi
oscillation survives in presence of dissipations and other non-ideal factors.
\end{abstract}

\section{Introduction}

Optomechanics and cavity quantum electrodynamics (QED) both study
light-matter interactions. While cavity QED involves light interacting with
nearly-resonant systems such as atoms \cite{dutra-2005, haroche-raimond-2006}%
, optomechanics treats highly off-resonant couplings between light and
mechanical objects \cite%
{wilson-Rae-kippenberg-093901-2007,marquardt-girvin-093902-2007,kippenberg-vahala-2008, aspelmeyer-schwab-2012, aspelmeyer-marquardt-2013}%
. The rapid developments in optomechanics towards the strong-coupling regime
make it possible to explore the quantum effects of single-photons on
macroscopic resonators \cite{brennecke-esslinger-2008, rabl-2011,
nunnenkamp-girvin-2011,stannigel-rabl-2012,komar-lukin-2013,qiao-nori-6302-2014,heikkila-sillanpaa-2014}%
, and hybrid systems consisting of optomechanics and cavity QED components
are expected to exhibit rich and novel features in the quantum regime \cite%
{hammerer-kimble-2009,restrepo-favero-2014,kyriienko-shelykh-2014,pirkkalainen-sillanpaa-6981-2015}.

As is well known, an excited two-level atom in a vacuum cavity undergoes
Rabi oscillation at a frequency proportional to the atom-photon coupling.
This is one excellent example illustrating how the cavity modifies the
optical properties of the atom, since in free space the atom would simply
decay exponentially. Adding a mechanical resonator to the atom-cavity system
via optomechanical coupling further changes the atomic behavior, even though
the mechanical resonator does not directly interact with the atom. In this
paper, we study the coherent dynamics of an excited two-level atom in a
vacuum optomechanical cavity. We find that as the atom-cavity Rabi splitting
is on resonance with the mechanical mode, the original Rabi oscillation is
\emph{sinusoidally modulated} by the optomechanical (light-mechanics)
interaction. This sinusoidal modulation of the Rabi oscillation is
non-trivial since it occurs in absence of any initial photons and phonons.
An analytic model in a \emph{three-dimensional} Hilbert subspace of the
hybrid system is developed to explain this intriguing phenomenon, and
numerical simulations of the density-matrix master equation are employed to
confirm the analytical analysis and to demonstrate that the result
qualitatively holds even as dissipations and other non-ideal factors are
taken into account.
\begin{figure}[tbh]
\vspace{-9cm} \centerline{\scalebox{1.0}{\includegraphics{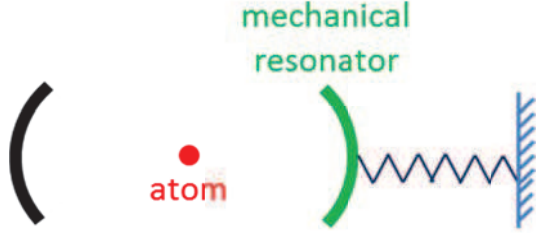}}}
\vspace{-16cm}
\caption{Schematic of a two-level atom inside an optomechanical cavity.}
\label{Fig1.fig}
\end{figure}

\section{Model and Hamiltonian}

We consider the setup in Fig. 1, where a two-level atom is placed in a
cavity with a movable end mirror as the mechanical resonator. The single
optical cavity mode is coupled to both the atom (via dipole-moment
interaction) and the mechanical resonator (via radiation pressure). The
hybrid atom-optomechanical system is described by the Hamiltonian ($\hbar =1$%
) \cite{restrepo-favero-2014}
\begin{equation}
\hat{H}=\omega _{c}\hat{c}^{\dagger }\hat{c}+\frac{\omega _{a}}{2}\hat{\sigma%
}_{z}+g_{ca}\left( \hat{\sigma}_{+}\hat{c}+\hat{\sigma}_{-}\hat{c}^{\dagger
}\right) +\omega _{m}\hat{b}^{\dagger }\hat{b}-g_{cm}\hat{c}^{\dagger }\hat{c%
}\left( \hat{b}+\hat{b}^{\dagger }\right) .  \label{H0.equ}
\end{equation}%
Here $\hat{c}$ and $\hat{b}$ are the annihilation operators for the optical
field (of frequency $\omega _{c}$) and the mechanical resonator (of
frequency $\omega _{m}$), respectively, $\hat{\sigma}_{z}=\left\vert
e\right\rangle _{a\ a}\left\langle e\right\vert -\left\vert g\right\rangle
_{a\;a}\left\langle g\right\vert $, $\hat{\sigma}_{+}=\hat{\sigma}%
_{-}^{\dagger }=\left\vert e\right\rangle _{a\ a}\left\langle g\right\vert $%
, with $\left\vert e\right\rangle _{a}$ ($\left\vert g\right\rangle _{a}$)
as the excited (ground) state of the atom, $\omega _{a}$ is the transition
frequency between the atomic states, and $g_{cm}$ ($g_{ca}$) characterizes
the strength of the light-mechanics (light-atom) interaction. In this
letter, we use subscripts \textquotedblleft $c$\textquotedblright ,
\textquotedblleft $a$\textquotedblright , \textquotedblleft $m$%
\textquotedblright\ to respectively denote \textquotedblleft optical
(cavity)\textquotedblright , \textquotedblleft atomic\textquotedblright\ and
\textquotedblleft mechanical\textquotedblright\ states or parameters. The
dimension of the hybrid-system Hilbert space is infinite. However, as we
will show later, if the system is initially in $\left\vert 0\right\rangle
_{c}\left\vert e\right\rangle _{a}\left\vert 0\right\rangle _{m}$, i.e.,
with the atom fully excited while the optical and the mechnical modes in the
ground states, then during the evolution the system is confined to a \emph{%
three-dimensional} Hilbert subspace, which allows for an analytical
description of the coherent dynamics.

We note that the photon number plus the population in the atomic excited
state is a constant of motion. Thus, starting from $\left\vert
0\right\rangle _{c}\left\vert e\right\rangle _{a}\left\vert 0\right\rangle
_{m},$ the quantum state of the hybrid system at any time $t$ stays in the
Hilbert subspace $\left\{ \left\vert 1\right\rangle _{c}\left\vert
g\right\rangle _{a},\ \left\vert 0\right\rangle _{c}\left\vert
e\right\rangle _{a}\right\} \otimes \mathcal{H}_{m}$,$\ $where $\left\vert
1\right\rangle _{c}$ is the one-photon Fock state, \textquotedblleft $%
\left\{ {}\right\} $\textquotedblright\ stands for the subspace spanned by
the basis vectors inside, and $\mathcal{H}_{m}$ represents the Hilbert space
of the mechanical mode. Assuming that the atom and the cavity are on
resonance, i.e., $\omega _{a}=\omega _{c}$, we construct the eigenstates of
the atom-cavity part of the Hamiltonian [$\hat{H}_{ca}=\omega _{c}\hat{c}%
^{\dagger }\hat{c}+\omega _{a}\hat{\sigma}_{z}/2+g_{ca}\left( \hat{\sigma}%
_{+}\hat{c}+\hat{\sigma}_{-}\hat{c}^{\dagger }\right) $] in the
two-dimensional subspace $\left\{ \left\vert 1\right\rangle _{c}\left\vert
g\right\rangle _{a},\ \left\vert 0\right\rangle _{c}\left\vert
e\right\rangle _{a}\right\} $ as%
\begin{equation}
\left\vert \pm \right\rangle _{ca}=\frac{\left\vert 1\right\rangle
_{c}\left\vert g\right\rangle _{a}\pm \ \left\vert 0\right\rangle
_{c}\left\vert e\right\rangle _{a}}{\sqrt{2}},  \label{new1.equ}
\end{equation}%
which correspond to the eigenvalues of $\omega _{a}/2\pm g_{ca}$. In this
new basis $\hat{H}_{ca}$ is diagonalized, i.e.,%
\begin{equation}
\hat{H}_{ca}=\frac{\omega _{a}}{2}+g_{ca}\hat{\sigma}_{z}^{\prime },
\label{Hca.equ}
\end{equation}%
with $\hat{\sigma}_{z}^{\prime }=\left\vert +\right\rangle _{ca\
ca}\left\langle +\right\vert -\left\vert -\right\rangle _{ca\
ca}\left\langle -\right\vert $. Moreover, since $\hat{c}^{\dagger }\hat{c}%
\left\vert \pm \right\rangle _{ca}=\left( \left\vert +\right\rangle
_{ca}+\left\vert -\right\rangle _{ca}\right) /2$, one has $\hat{c}^{\dagger }%
\hat{c}=\left( 1+\hat{\sigma}_{+}^{\prime }+\hat{\sigma}_{-}^{\prime
}\right) /2$, where $\hat{\sigma}_{+}^{\prime }=\hat{\sigma}_{-}^{\prime
\dagger }=\left\vert +\right\rangle _{ca\ ca}\left\langle -\right\vert $.
Thus in the subspace\ $\left\{ \left\vert 1\right\rangle _{c}\left\vert
g\right\rangle _{a},\ \left\vert 0\right\rangle _{c}\left\vert
e\right\rangle _{a}\right\} \otimes \mathcal{H}_{m}$, the total Hamiltonian (%
\ref{H0.equ}) becomes\ $\hat{H}=\omega _{a}/2+g_{ca}\hat{\sigma}_{z}^{\prime
}+\omega _{m}\hat{b}^{\dagger }\hat{b}-g_{cm}\left( 1+\hat{\sigma}%
_{+}^{\prime }+\hat{\sigma}_{-}^{\prime }\right) \left( \hat{b}+\hat{b}%
^{\dagger }\right) /2.$ By introducing the displaced phonon operator\ $\hat{b%
}^{\prime }=\hat{b}-g_{cm}/\left( 2\omega _{m}\right) $ the Hamiltonian is
re-written as%
\[
\hat{H}=g_{ca}\hat{\sigma}_{z}^{\prime }+\omega _{m}\hat{b}^{\prime \dagger }%
\hat{b}^{\prime }-\frac{g_{cm}}{2}\left( \hat{\sigma}_{+}^{\prime }+\hat{%
\sigma}_{-}^{\prime }\right) \left( \hat{b}^{\prime }+\hat{b}^{\prime
\dagger }\right) -\frac{g_{cm}^{2}}{2\omega _{m}}\left( \hat{\sigma}%
_{+}^{\prime }+\hat{\sigma}_{-}^{\prime }\right) ,
\]%
where a constant energy term\ $\omega _{a}/2-g_{cm}^{2}/\left( 4\omega
_{m}\right) $ has been discarded. Further assuming that\ $g_{cm}\ll \omega
_{m}$ and $\omega _{m}\approx 2g_{ca}$, we neglect the term proportional to $%
g_{cm}^{2}$ and make the rotating wave approximation in the above
Hamiltonian to obtain $\hat{H}\approx \hat{H}_{eff}$, with%
\begin{equation}
\hat{H}_{eff}=g_{ca}\hat{\sigma}_{z}^{\prime }+\omega _{m}\hat{b}^{\prime
\dagger }\hat{b}^{\prime }-\frac{g_{cm}}{2}\left( \hat{\sigma}_{+}^{\prime }%
\hat{b}^{\prime }+\hat{\sigma}_{-}^{\prime }\hat{b}^{\prime \dagger }\right)
.  \label{H3.equ}
\end{equation}%
This Hamiltonian describes a two-level \textquotedblleft
polariton\textquotedblright\ coupled to a bosonic mode, analogous to the
original atom-cavity Hamiltonian $\hat{H}_{ca}$.

\section{Sinusoidally Modulated Rabi Oscillation}

In the previous section, we derived an effective Hamiltonian $\hat{H}_{eff}$
in (\ref{H3.equ}) for the hybrid system where initially the atom is fully
excited and both the optical and the mechanical modes are in their ground
states. At first glance, since $\hat{H}_{eff}$ is a J-C type Hamiltonian, it
might suggest a simple Rabi oscillation of frequency $g_{cm}$ between \emph{%
two} basis vectors, analogous to that in a regular atom-cavity system (i.e.,
without the mechanical resonator). This would be true if the hybrid system
were initially in $\left\vert +\right\rangle _{ca}\left\vert 0\right\rangle
_{m^{\prime }}$, with $\left\vert 0\right\rangle _{m^{\prime }}$ being the
ground state of the displaced mechanical mode $\hat{b}^{\prime }$. However,
the initial state of the system is not $\left\vert +\right\rangle
_{ca}\left\vert 0\right\rangle _{m^{\prime }}$, but rather $\left\vert
0\right\rangle _{c}\left\vert e\right\rangle _{a}\left\vert 0\right\rangle
_{m}$. As will become clear below, starting from $\left\vert 0\right\rangle
_{c}\left\vert e\right\rangle _{a}\left\vert 0\right\rangle _{m}$, the
evolution of the hybrid system involves \emph{three} (rather than two) basis
vectors, which in turn leads to a non-trivial modulation of the original
atom-cavity Rabi oscillation by the optomechanical coupling.

We first take a deeper look at the initial state $\left\vert 0\right\rangle
_{c}\left\vert e\right\rangle _{a}\left\vert 0\right\rangle _{m}$. Although
rigorously the ground state $\left\vert 0\right\rangle _{m}$ of the original
mechanical mode $\hat{b}$ is the coherent state of the displaced mode $\hat{b%
}^{\prime }$ with the eigenvalue $-g_{cm}/\left( 2\omega _{m}\right) $, we
can well approximate $\left\vert 0\right\rangle _{m}$ $\approx $ $\left\vert
0\right\rangle _{m^{\prime }}$ since it has been assumed that $g_{cm}\ll
\omega _{m}$. According to (\ref{new1.equ}), $\left\vert 0\right\rangle
_{c}\left\vert e\right\rangle _{a}$ is a superposition of $\left\vert
+\right\rangle _{ca}$ and $\left\vert -\right\rangle _{ca}$, and thus the
initial state $\left\vert 0\right\rangle _{c}\left\vert e\right\rangle
_{a}\left\vert 0\right\rangle _{m}\approx \left\vert 0\right\rangle
_{c}\left\vert e\right\rangle _{a}\left\vert 0\right\rangle _{m^{\prime }}$
is a superposition of $\left\vert +\right\rangle _{ca}\left\vert
0\right\rangle _{m^{\prime }}$ and $\left\vert -\right\rangle
_{ca}\left\vert 0\right\rangle _{m^{\prime }}$. Further noting that $%
\left\vert -\right\rangle _{ca}\left\vert 0\right\rangle _{m^{\prime }}$ is
an eigenstate of $\hat{H}_{eff}$, and $\hat{H}_{eff}$ couples $\left\vert
+\right\rangle _{ca}\left\vert 0\right\rangle _{m^{\prime }}$ only to $%
\left\vert -\right\rangle _{ca}\left\vert 1\right\rangle _{m^{\prime }}$ and
vice versa, we conclude that the coherent dynamics of the system is
constrained to the three-dimensional Hilbert subspace of
\begin{equation}
\mathcal{H}^{\left( 3\right) }=\left\{ \left\vert -\right\rangle
_{ca}\left\vert 0\right\rangle _{m^{\prime }},\ \left\vert +\right\rangle
_{ca}\left\vert 0\right\rangle _{m^{\prime }},\ \left\vert -\right\rangle
_{ca}\left\vert 1\right\rangle _{m^{\prime }}\right\} ,  \label{Hil3.equ}
\end{equation}%
which allows us to obtain an analytic solution for the evolution of the
hybrid system.

To this end, we diagonalize $\hat{H}_{eff}$ in the subspace $\mathcal{H}%
^{\left( 3\right) }$%
\begin{equation}
\hat{H}_{eff}=-g_{ca}\left\vert \tilde{0}\right\rangle \left\langle \tilde{0}%
\right\vert +\Omega _{+}\left\vert \tilde{+}\right\rangle \left\langle
\tilde{+}\right\vert +\Omega _{-}\left\vert \tilde{-}\right\rangle
\left\langle \tilde{-}\right\vert ,  \label{Heff2.equ}
\end{equation}%
where%
\begin{eqnarray}
\left\vert \tilde{0}\right\rangle  &=&\left\vert -\right\rangle
_{ca}\left\vert 0\right\rangle _{m^{\prime }},\ \left\vert \tilde{+}%
\right\rangle =\mathcal{N}\left( \left\vert -\right\rangle _{ca}\left\vert
1\right\rangle _{m^{\prime }}-\eta \left\vert +\right\rangle _{ca}\left\vert
0\right\rangle _{m^{\prime }}\right) ,  \nonumber \\
\ \left\vert \tilde{-}\right\rangle  &=&\mathcal{N}\left( \eta \left\vert
-\right\rangle _{ca}\left\vert 1\right\rangle _{m^{\prime }}+\left\vert
+\right\rangle _{ca}\left\vert 0\right\rangle _{m^{\prime }}\right) ,
\label{new2.equ}
\end{eqnarray}%
and $\Omega _{\pm }=\left( \omega _{m}\pm \Omega _{g}\right) /2,\ \Omega
_{g}=\sqrt{\left( \omega _{m}-2g_{ca}\right) ^{2}+g_{cm}^{2}},$ $\eta =\left[
\Omega _{g}-\left( \omega _{m}-2g_{ca}\right) \right] /g_{cm},\ \mathcal{N}%
=\left( 1+\eta ^{2}\right) ^{-1/2}$. With the initial state of the system
decomposed [by inverting (\ref{new1.equ}) and (\ref{new2.equ})] in terms of
the three eigenvectors of $\hat{H}_{eff}$ as%
\[
\left\vert \psi \left( t=0\right) \right\rangle =\left\vert 0\right\rangle
_{c}\left\vert e\right\rangle _{a}\left\vert 0\right\rangle _{m^{\prime }}=-%
\frac{\mathcal{N}\left( \eta \left\vert \tilde{+}\right\rangle -\left\vert
\tilde{-}\right\rangle \right) +\left\vert \tilde{0}\right\rangle }{\sqrt{2}}%
,
\]%
we get the quantum state at any time $t$%
\[
\left\vert \psi \left( t\right) \right\rangle =-\frac{\mathcal{N}}{\sqrt{2}}%
\eta e^{-i\frac{\omega _{m}+\Omega _{g}}{2}t}\left\vert \tilde{+}%
\right\rangle +\frac{\mathcal{N}}{\sqrt{2}}e^{-i\frac{\omega _{m}-\Omega _{g}%
}{2}t}\left\vert \tilde{-}\right\rangle -\frac{1}{\sqrt{2}}%
e^{ig_{ca}t}\left\vert \tilde{0}\right\rangle .
\]

To calculate the probability for the atom to stay in the excited state, we
convert $\left\vert \tilde{\pm}\right\rangle $ and $\left\vert \tilde{0}%
\right\rangle $ in $\left\vert \psi \left( t\right) \right\rangle $ back to
the \textquotedblleft original\textquotedblright\ basis vectors of $%
\left\vert 0\right\rangle _{c}\left\vert e\right\rangle _{a}\left\vert
0\right\rangle _{m^{\prime }}$, $\left\vert 1\right\rangle _{c}\left\vert
g\right\rangle _{a}\left\vert 0\right\rangle _{m^{\prime }}$, $\left\vert
0\right\rangle _{c}\left\vert e\right\rangle _{a}\left\vert 1\right\rangle
_{m^{\prime }}$ and $\left\vert 1\right\rangle _{c}\left\vert g\right\rangle
_{a}\left\vert 1\right\rangle _{m^{\prime }}$, which can be done by simply
substituting (\ref{new1.equ}) into (\ref{new2.equ}). Here we do not present
the detailed expression of $\left\vert \psi \left( t\right) \right\rangle $
in the original basis, but only write down its probability amplitudes $%
\mathcal{A}_{0e1}$ for $\left\vert 0\right\rangle _{c}\left\vert
e\right\rangle _{a}\left\vert 1\right\rangle _{m^{\prime }}$ and $\mathcal{A}%
_{0e0}$ for $\left\vert 0\right\rangle _{c}\left\vert e\right\rangle
_{a}\left\vert 0\right\rangle _{m^{\prime }}$:
\begin{eqnarray*}
\mathcal{A}_{0e1} &=&-i\mathcal{N}^{2}\eta e^{-i\frac{\omega _{m}}{2}t}\sin
\left( \frac{\Omega _{g}}{2}t\right) ,\  \\
\mathcal{A}_{0e0} &=&\frac{\mathcal{N}^{2}e^{-i\frac{\omega _{m}}{2}t}\left(
\eta ^{2}e^{-i\frac{\Omega _{g}}{2}t}+e^{i\frac{\Omega _{g}}{2}t}\right)
+e^{ig_{oa}t}}{2}.
\end{eqnarray*}%
The probability $P_{e}$ to find the atom in the excited state is given by $%
\left\vert \mathcal{A}_{0e1}\right\vert ^{2}+\left\vert \mathcal{A}%
_{0e0}\right\vert ^{2}$. In terms of the more widely used poulation
inversion $\Delta P=P_{e}-P_{g}$, one has%
\begin{equation}
\Delta P=\cos \left( \Omega _{ca}t\right) \cos \frac{\Omega _{g}t}{2}+\xi
\sin \left( \Omega _{ca}t\right) \sin \frac{\Omega _{g}t}{2},  \label{DP.equ}
\end{equation}%
where
\[
\Omega _{ca}=\frac{\omega _{m}}{2}+g_{ca},\ \xi =\frac{1-\eta ^{2}}{1+\eta
^{2}},
\]%
and $P_{g}=1-P_{e}$ is the probability for the atom to be in the ground
state. Eq. (\ref{DP.equ}) is drastically different from the standard Rabi
oscillation $\Delta P_{ca}=\cos \left( 2g_{ca}t\right) $ in the atom-cavity
system governed by Hamiltonian $\hat{H}_{ca}$ [(\ref{Hca.equ})].
Particularly, if the atom-cavity Rabi splitting is on resonance with the
mechanical mode, i.e., $\omega _{m}=2g_{ca}$, we can simplify (\ref{DP.equ})
into%
\begin{equation}
\Delta P=\cos \left( 2g_{ca}t\right) \cos \left( \frac{g_{cm}}{2}t\right) ,
\label{DP1.equ}
\end{equation}%
which shows that \emph{the optomechanical coupling modulates the original
atom-cavity Rabi oscilation with a cosine envelope function}. This
sinusoidally modulated Rabi oscillation is our central result, and its
derivation has involved a fairly large amount of mathematics, so in next
paragraph we will provide a more physically intuitive argument.

To do this, we analyze the action of $\hat{H}_{eff}$ in (\ref{H3.equ}) on
the three dimensional Hilbert subspace $\mathcal{H}^{\left( 3\right) }$ in (%
\ref{Hil3.equ}). We note that $\hat{H}_{eff}$ induces two distinct
oscillatory transitions: on a short time scale of $g_{ca}^{-1}$, the
atom-cavity interaction ($g_{ca}\hat{\sigma}_{z}^{\prime }$) yields the Rabi
oscillation between $\left\vert 0\right\rangle _{c}\left\vert e\right\rangle
_{a}\left\vert 0\right\rangle _{m^{\prime }}=\left( \left\vert
+\right\rangle _{ca}-\left\vert -\right\rangle _{ca}\right) \left\vert
0\right\rangle _{m^{\prime }}/\sqrt{2}$ and $\left\vert 1\right\rangle
_{c}\left\vert g\right\rangle _{a}\left\vert 0\right\rangle _{m^{\prime
}}=\left( \left\vert +\right\rangle _{ca}+\left\vert -\right\rangle
_{ca}\right) \left\vert 0\right\rangle _{m^{\prime }}/\sqrt{2}$, or
equivalently, the oscillation of the dynamical phase difference between the $%
\left\vert +\right\rangle _{ca}\left\vert 0\right\rangle _{m^{\prime }}$ and
$\left\vert -\right\rangle _{ca}\left\vert 0\right\rangle _{m^{\prime }}$
components; on a long time scale of $g_{cm}^{-1}$, the optomechanical
coupling [$-g_{cm}\left( \hat{\sigma}_{+}^{^{\prime }}\hat{b}^{^{\prime }}+%
\hat{\sigma}_{-}^{^{\prime }}\hat{b}^{^{\prime }\dagger }\right) /2$] causes
the \textquotedblleft polariton-phonon\textquotedblright\ oscillation
between $\left\vert +\right\rangle _{ca}\left\vert 0\right\rangle
_{m^{\prime }}$ and $\left\vert -\right\rangle _{ca}\left\vert
1\right\rangle _{m^{\prime }}$. At $t=0$, the system is in $\left\vert
0\right\rangle _{c}\left\vert e\right\rangle _{a}\left\vert 0\right\rangle
_{m^{\prime }}=\left( \left\vert +\right\rangle _{ca}-\left\vert
-\right\rangle _{ca}\right) \left\vert 0\right\rangle _{m^{\prime }}/\sqrt{2}
$, and the atom-cavity interaction generates the Rabi oscillation on the
short time scale. After a duration of $\pi /g_{cm}$, $\left\vert
+\right\rangle _{ca}\left\vert 0\right\rangle _{m^{\prime }}$ evolves into $%
\left\vert -\right\rangle _{ca}\left\vert 1\right\rangle _{m^{^{\prime }}}$
due to the optomechanical coupling, and the state of the system becomes $%
\left\vert -\right\rangle _{ca}\left( \left\vert 1\right\rangle _{m^{\prime
}}+e^{i\phi }\left\vert 0\right\rangle _{m^{\prime }}\right) /\sqrt{2}$,
where $\phi $ is a phase difference between the two Fock states of the
(displaced) mechanical mode. Around this time ($t\approx \pi /g_{cm}$) there
is no atom-cavity Rabi oscillation since the $\left\vert +\right\rangle
_{ca}\left\vert 0\right\rangle _{m^{\prime }}$ component dose not exist in
the system state. After another duration of $\pi /g_{cm}$, i.e., at $t=2\pi
/g_{cm}$, $\left\vert -\right\rangle _{ca}\left\vert 1\right\rangle
_{m^{\prime }}$ evolves back to $\left\vert +\right\rangle _{ca}\left\vert
0\right\rangle _{m^{\prime }}$, and the atom-cavity Rabi oscillation
re-appears. For other $t$ $\in \left( 0,2\pi /g_{cm}\right) $, partial Rabi
oscillations (i.e., without completely reaching the ground or the excited
states of the atom) occur. This pattern repeats in every period of $2\pi
/g_{cm}$, leading to the periodically-modulated Rabi oscillation in (\ref%
{DP1.equ}). We notice that the envelope function in (\ref{DP1.equ}) has a
period of $4\pi /g_{cm}$, but the Rabi oscillation on the short time scale
does not distinguish positive and negative values of the envelope function,
and thus the \textquotedblleft modulation period\textquotedblright\ is
indeed $2\pi /g_{cm}$, agreeing with our analysis above. We further remark
that the involvement of three (rather than two) basis vectors in $\mathcal{H}%
^{\left( 3\right) }$ is crucial to the sinusoidally modulated Rabi
oscillation, because it is the optomechanical transition between $\left\vert
+\right\rangle _{ca}\left\vert 0\right\rangle _{m^{^{\prime }}}$ and $%
\left\vert -\right\rangle _{ca}\left\vert 1\right\rangle _{m^{^{\prime }}}$
that gives rise to the modulation of the regular atom-cavity Rabi
oscillation in the subspace of $\left\{ \left\vert +\right\rangle
_{ca}\left\vert 0\right\rangle _{m^{^{\prime }}},\left\vert -\right\rangle
_{ca}\left\vert 0\right\rangle _{m^{^{\prime }}}\right\} $.

\section{Numerical Simulations and Non-Ideal Situations}

\begin{figure}[tbh]
\vspace*{-1cm}\hspace*{0cm} \centerline{\scalebox{0.5}{%
\includegraphics{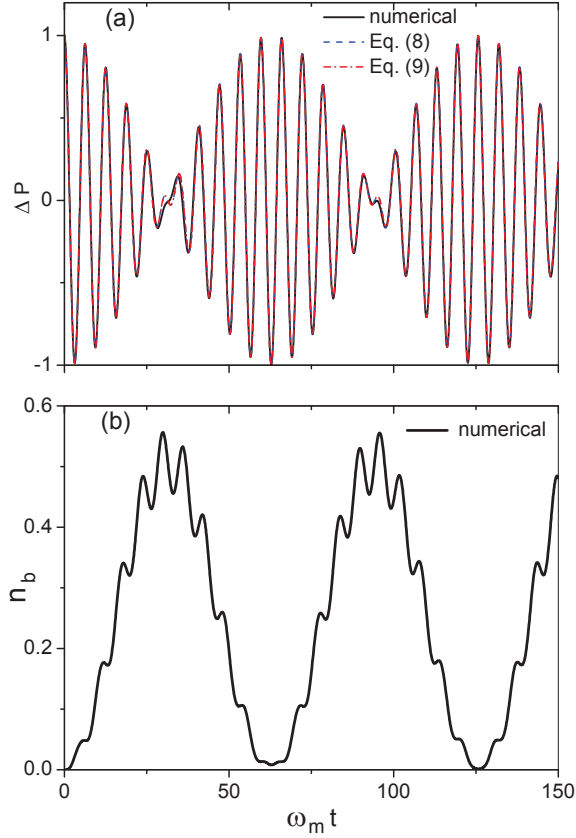}}} \vspace*{-3cm}
\caption{Temporal evolution of (a) the atomic population inversion, and (b)
the mean phonon number, with $g_{cm}=0.1\protect\omega _{m}$, $\protect%
\omega _{a}=\protect\omega _{c}$, $g_{ca}=0.5\protect\omega _{m}$, and no
thermal effects. The three curves in (a) are almost indistinguishable from
each other, and the sinusoidally modulated Rabi oscillation is nearly
perfect.}
\label{Fig2.fig}
\end{figure}

\begin{figure}[tbh]
\vspace*{-1cm}\hspace*{0cm} \centerline{\scalebox{0.5}{%
\includegraphics{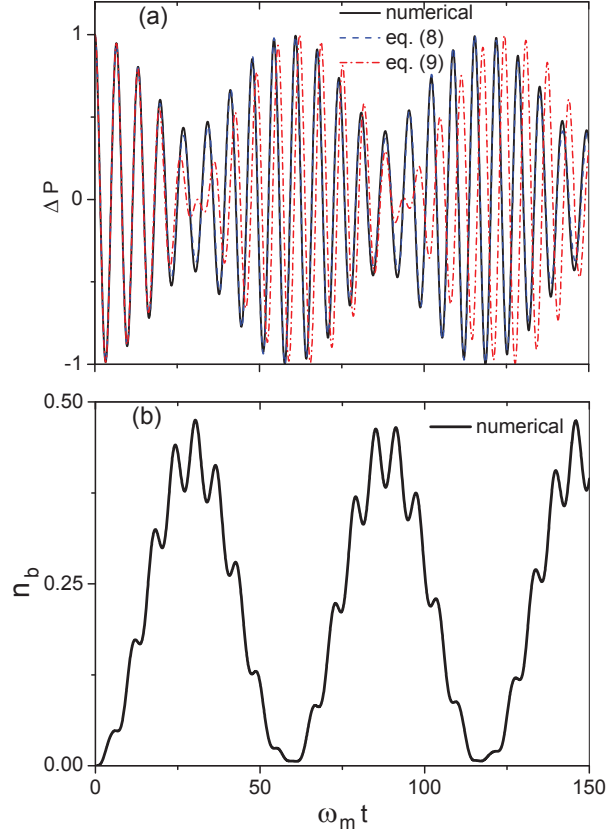}}} \vspace*{-3cm}
\caption{Temporal evolution of (a) the atomic population inversion, and (b)
the mean phonon number, with $g_{cm}=0.1\protect\omega _{m}$, $\protect%
\omega _{a}=\protect\omega _{c}$, $g_{ca}=0.48\protect\omega _{m}$, and no
thermal effects. In (a), the curve for Eq. (\protect\ref{DP.equ})
essentially coincides with the numerical one, which shows a non-ideal
modulated Rabi oscillation.}
\label{Fig3.fig}
\end{figure}

\begin{figure}[tbh]
\vspace*{-4cm}\hspace*{0cm} \centerline{\scalebox{0.5}{%
\includegraphics{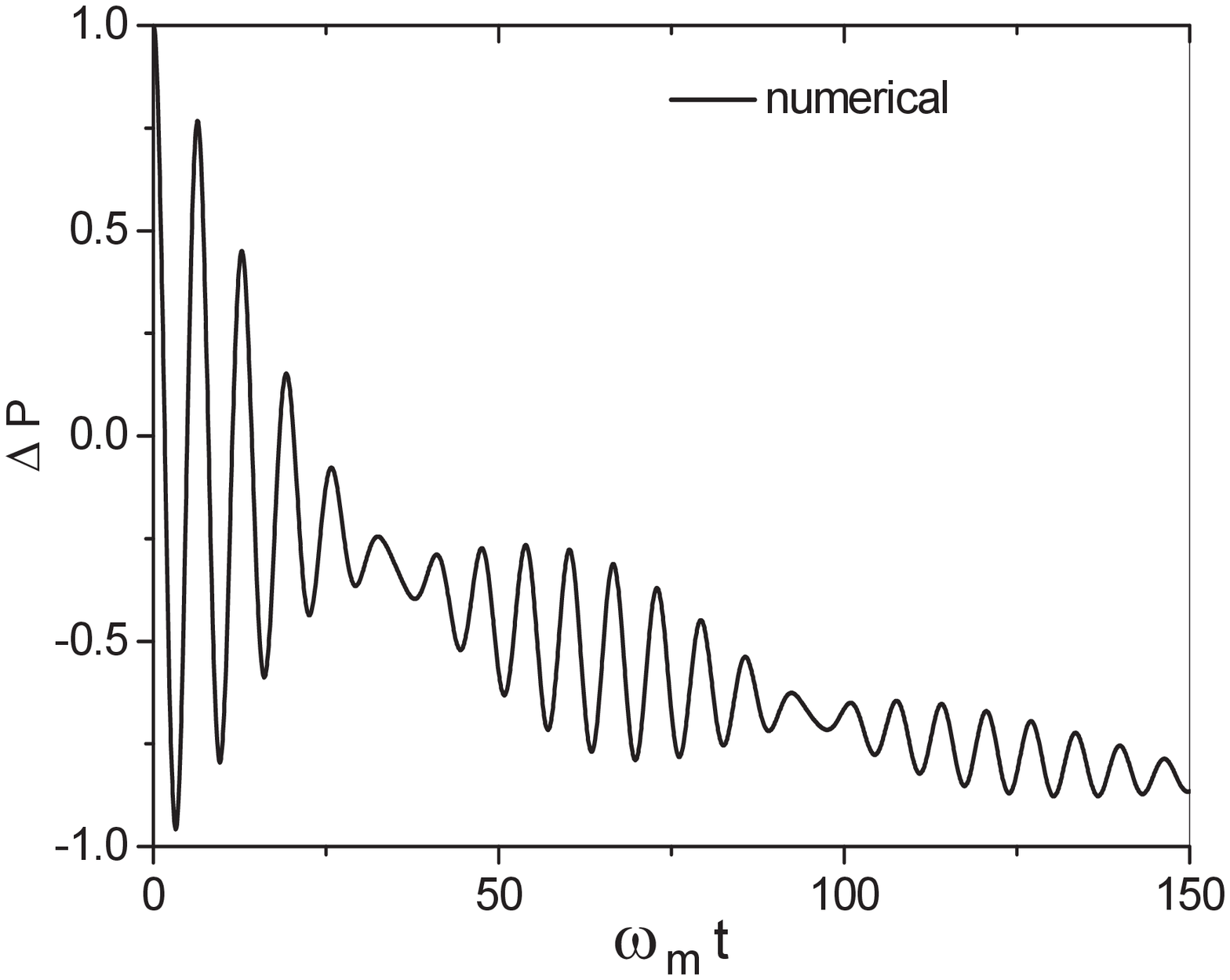}}} \vspace*{-4cm}
\caption{Modulated Rabi oscillation in presence of thermal effects. The
parameters are $g_{cm}=0.1\protect\omega _{m}$, $\protect\omega _{a}=\protect%
\omega _{c}-0.01\protect\omega _{m}$, $g_{ca}=0.49\protect\omega _{m}$, $%
\protect\kappa =0.02\protect\omega _{m}$, $\protect\gamma =0.005\protect%
\omega _{m}$, $\protect\mu =2\times 10^{-4}\protect\omega _{m}$, $n_{th}=10$%
, and the initial thermal phonon number of the mechanical resonator is $%
n_{th}^{\left( 0\right) }=0.5$.}
\label{Fig4.fig}
\end{figure}
Eq. (\ref{DP1.equ}), which illustrates the sinusoidal modulation of the Rabi
oscillation in the atom-optomechanical system, has been derived with the
following assumptions: no dissipations, $\omega _{c}=\omega _{a}$, $\omega
_{m}=2g_{ca}$, $g_{cm}\ll \omega _{m}$, and deviations from these ideal
conditions are expected to affect the dynamics. For example, if $\omega _{m}$
is slightly detuned from $2g_{ca}$ (and thus $\xi \neq 0$), then $\Delta P$
is determined by Eq. (\ref{DP.equ}) [rather than Eq. (\ref{DP1.equ})], where
the second term on the right-hand side perturbs the \textquotedblleft
ideal\textquotedblright\ dynamics manifested in the first term. To more
comprehensively understand the robustness of the modulated Rabi oscillation,
we resort to numerical simulations of the density-matrix ($\hat{\rho}$)
master equation \cite{scully-zubairy-1997} for the hybrid system%
\begin{equation}
\frac{d\hat{\rho}}{dt}=\frac{1}{i}\left[ \hat{H},\hat{\rho}\right] -\kappa L%
\left[ \hat{a}\right] \hat{\rho}-\gamma L\left[ \hat{\sigma}_{-}\right] \hat{%
\rho}-\left( n_{th}+1\right) \mu L\left[ \hat{b}\right] \hat{\rho}-n_{th}\mu
L\left[ \hat{b}^{\dagger }\right] \hat{\rho},  \label{DME.equ}
\end{equation}%
where $\hat{H}$ is the unapproximated Hamiltonian in (\ref{H0.equ}), $L\left[
\hat{o}\right] \hat{\rho}=\hat{o}^{\dagger }\hat{o}\hat{\rho}/2-\hat{o}\hat{%
\rho}\hat{o}^{\dagger }+\hat{\rho}\hat{o}^{\dagger }\hat{o}/2$, $\kappa $, $%
\gamma $, $\mu $ are the decay rates of the optical cavity, the atom, and
the mechanical resonator, respectively, and $n_{th}$ is the thermal
occupation of the mechanical heat-bath mode at frequency $\omega _{m}$. It
has also been assumed that both the optical and the atomic heat baths have
zero thermal occupation at $\omega _{c}$ and $\omega _{a}$, which is a good
approximation at room temperature for $\omega _{c}$ and $\omega _{a}$ within
the visible-light frequency range. The population inversion of the atom is
calculated from the density matrix via $\Delta P=Tr\left[ \hat{\rho}\hat{%
\sigma}_{z}\right] $, with \textquotedblleft $Tr$\textquotedblright\ being
the trace of operators.

We first check the validity of our analytic model for the sinusoidally
modulated Rabi oscillation. In each of Figs. 2a and 3a we plot three curves:
Eqs. (\ref{DP.equ}) [blue dash] and (\ref{DP1.equ}) [red dash-dot], and the
numerical simulation of the master equation (\ref{DME.equ}) [black solid]
with $\rho \left( t=0\right) =\left\vert 0\right\rangle _{c}\left\vert
e\right\rangle _{a}\left\vert 0\right\rangle _{m\;m}\left\langle
0\right\vert _{\;a}\left\langle e\right\vert _{\;c}\left\langle 0\right\vert
$ and $\kappa =\gamma =\mu =0$. In Fig. 2a, we take the parameters to be $%
\omega _{a}=\omega _{c}$, $g_{ca}=0.5\omega _{m}$, $g_{cm}=0.1\omega _{m}$,
for which Eq. (\ref{DP1.equ}) is equivalent to the more general Eq. (\ref%
{DP.equ}). We find that the two analytic curves well fit the numerical one,
and the system undergoes a perfect sinusoidally-modulated Rabi oscillation.
In Fig. 3a, we set $\omega _{a}=\omega _{c}$, $g_{ca}=0.48\omega _{m}$, $%
g_{cm}=0.1\omega _{m}$. For these parameters, Eq. (\ref{DP1.equ}) is no
longer valid since $\omega _{m}\neq 2g_{ca}$, but Eq. (\ref{DP.equ}) still
holds. In this case, Eq. (\ref{DP.equ}) agrees well with the numerical
result, and the modulation of the Rabi oscillation is not ideally
sinusoidal. Fig. 2a and Fig. 3a confirm that our analytic model correctly
describes the coherent dynamics of the hybrid system within the proper
parameter range.

In Figs. 2b and 3b, we plot the numerical results for the temporal evolution
of the phonon number $n_{b}=Tr\left[ \hat{\rho}\hat{b}^{\dagger }\hat{b}%
\right] $, with the same parameters as in Figs. 2a and 3a, respectively.
According to the analysis in the last paragraph of Sec. III, the
Rabi-oscillation amplitude is maximum (minimum) when the phonon number is
minimum (maximum). Comparing Fig. 2a and Fig. 2b, as well as Figs. 3a and
3b, immediately confirms this conclusion, and thus provides further support
for our analytic model in the previous section.

Next we investigate the situation where dissipations are present by adopting
$\kappa =0.02\omega _{m}$, $\gamma =0.005\omega _{m}$, $\mu =2\times
10^{-4}\omega _{m}$, $n_{th}=10$. We also allow that the other parameters do
not rigorously meet (but not deviate too much from) the requirements for Eq.
(\ref{DP1.equ}), with $\omega _{a}=\omega _{c}-0.1\omega _{m}$, $%
g_{ca}=0.49\omega _{m}$, and $g_{cm}=0.1\omega _{m}$. Moreover, since the
mechanical mode can at best be \emph{pre-cooled} to a small but finite mean
occupation number $n_{th}^{\left( 0\right) }<n_{th}$, we set the mechanical
resonator to be initially in a thermal state (rather than the ground state)
of $n_{th}^{\left( 0\right) }=0.5$. The numerical solution for $\Delta P$,
as plotted in Fig. 4, unambiguously demonstrates that \emph{the modulated
Rabi oscillation survives} even though it is siginficantly degraded from the
ideal one illustrated in Eq. (\ref{DP1.equ}) [or in Fig. 2a].

\section{Summary}

In summary, the quantum behavior of an excited two-level atom in a vacuum
optomechanical cavity can be qualitatively modified by the mechanical
resonator. Particularly, if the atom-cavity Rabi splitting is on resonance
with the mechanical mode, then the standard Rabi oscillation of the atom is
sinusoidally modulated. The modulation originates from the
``polariton-phonon'' transition caused by the coupling between the optical
field and the mechanical resonator. We explained this phenomenon with an
analytic model in a three-dimensional Hilbert subspace of the hybrid system,
which was further confirmed by numerical simulations of the density-matrix
master equation. The modulated Rabi oscillation was found to be reasonably
tolerant on thermal effects and non-idealized parameters.

\section{Acknowledgement}

We gratefully acknowledge financial supports from the National Natural
Science Foundation of China (No. 11375081, 11347026), the Shandong
Provincial Natural Science Foundation (No. ZR2013AL007, ZR2013AM012), and
the Start-up Fund of Liaocheng University.

$\;$

\end{document}